\documentclass[11pt]{article}
\usepackage{amsmath,amssymb,color,epsfig,cite}
\usepackage{graphicx}
\usepackage{subfigure}
\usepackage{setspace}

\usepackage{amsfonts}

\newcommand{\hoch}[1]{$\, ^{#1}$}

\makeatletter
\@addtoreset{equation}{section}
\makeatother

\newcommand{\be}{\begin{equation}}
\newcommand{\ee}{\end{equation}}
\newcommand{\bea}{\setlength\arraycolsep{2pt} \begin{eqnarray}}
\newcommand{\eea}{\end{eqnarray}}
\newcommand{\nn}{\nonumber}

\def\ft#1#2{{\textstyle{\frac{\scriptstyle #1}{\scriptstyle #2} } }}
\def\fft#1#2{{\frac{#1}{#2}}}

\def\0{{\sst{(0)}}}
\def\1{{\sst{(1)}}}
\def\2{{\sst{(2)}}}
\def\3{{\sst{(3)}}}
\def\4{{\sst{(4)}}}
\def\5{{\sst{(5)}}}
\def\6{{\sst{(6)}}}
\def\7{{\sst{(7)}}}
\def\8{{\sst{(8)}}}
\def\sst#1{{\scriptscriptstyle #1}}

\def\del{{\partial}}

\begin{document}

\begin{flushright}
\hfill{MI-TH-1762}

\end{flushright}

\begin{center}
{\large {\bf Holographic Heat Current as Noether Current}}

\vspace{10pt}
Hai-Shan Liu\hoch{1}, H. L\"u\hoch{2} and C.N. Pope\hoch{3,4}

\vspace{10pt}

\hoch{1}{\it Institute for Advanced Physics \& Mathematics,\\
Zhejiang University of Technology, Hangzhou 310023, China}

\vspace{10pt}

\hoch{2}{\it Center for Advanced Quantum Studies, Department of Physics,\\
Beijing Normal University, Beijing 100875, China}

\vspace{10pt}

\hoch{3}{\it George P. \& Cynthia Woods Mitchell  Institute
for Fundamental Physics and Astronomy,\\
Texas A\&M University, College Station, TX 77843, USA}

\vspace{10pt}

\hoch{4}{\it DAMTP, Centre for Mathematical Sciences,
 Cambridge University,\\  Wilberforce Road, Cambridge CB3 OWA, UK}

\vspace{30pt}

\underline{ABSTRACT}

\end{center}

We employ the Noether procedure to derive a general formula for the
radially conserved heat current in AdS planar black holes with
certain transverse and traceless perturbations, for a general class of
gravity theories.  For Einstein gravity, the general higher-order
Lovelock gravities and also a class of Horndeski gravities, we derive
the boundary stress tensor and show that the resulting boundary heat current
matches precisely the bulk Noether current.

\vfill {\footnotesize Emails: hsliu.zju@gmail.com \ \ \ mrhonglu@gmail.com\ \ \ pope@physics.tamu.edu}

\thispagestyle{empty}

\pagebreak

\tableofcontents
\addtocontents{toc}{\protect\setcounter{tocdepth}{2}}



\section{Introduction}

Recently, gauge/gravity duality has been used to understand various
phenomena for strongly coupled systems in condensed matter
physics \cite{ggd1,ggd2,ggd3,ggd4}. In particular, holographic gravity models
with momentum relaxation have attracted much attention. These
models can serve as a realistic description of materials with
impurities \cite{lat1,lat2,lat3,mg1,lat4,mg2,lat5,lat8,EMDC,DC1,DC2,lat6,lat7,mg3,gsge,DC3,DC4,DC5,gels2,
gels1,Cremonini:2016avj,donoshgr,Jiang:2017imk,Baggioli:2017ojd}.

With the help of gauge/gravity duality, one can calculate the transport
coefficients in these strongly coupled systems by analysing the linear
response to a small perturbation around a black hole background which
describes an equilibrium state. Among these transport coefficients,
much effort has been directed to calculating the AC conductivity,
mostly involving the use of numerical methods. However, ways to calculate
the DC conductivity analytically have also been developed, based on
the ``membrane paradigm''\cite{DChong}. The key step involves constructing
a radially conserved current \cite{mg2}, which provides an analytical
relation between the holographic boundary information and the black hole
horizon data.

With the same philosophy, one can obtain the holographic thermal and
thermoelectric conductivities. This was developed in \cite{DC2} for
the case of Einstein-Maxwell-Dilaton (EMD) gravity, by manipulating the
equations of motion to construct a radially conserved bulk heat current.
However, unlike the electric current which arises naturally from the
Maxwell equation, the bulk heat current is considerably more subtle to
calculate, since it makes use of both the Einstein and Maxwell equations.
The problem can be further exacerbated in higher-order gravity theories.
Matching the bulk current to the boundary heat current, derived from the
boundary stress tensor, can also be quite involved. A general procedure
for constructing the holographic electric and heat currents for higher
derivative gravity by using dimension reduction was proposed in
\cite{donoshgr}.  However,  a general and simple
formula for this bulk current is lacking in literature; the known examples
were obtained for certain specific theories. The first example was given
by Donos and Gauntlett \cite{DC2} for the EMD theory. Another notable
example is the holographic heat current in Einstein-Gauss-Bonnet
gravity \cite{gsge,donoshgr}.

The purpose of this paper is to present a general formula for such
radially conserved currents associated with certain transverse and
traceless (TT) perturbation of the AdS planar black hole. We follow
closely the procedure described by Wald \cite{wald1,wald2}, and show
that the radially conserved current is simply the Noether current
associated with the time-like Killing vector.  We then show that the
Noether current indeed matches the boundary heat current for
AdS planar black holes in Einstein gravity, general Lovelock
gravities \cite{ll} and also a class of Horndeski gravities
\cite{Horndeski:1974wa}.

The paper is organized as follows.  In section 2, we consider a
general class of gravity theories and use the Noether procedure to
derive a general formula for the radially conserved current.  In section 3,
we consider the simple example of Einstein-Maxwell-Axion (EMA) theory,
for which we derive the boundary stress tensor and show that the bulk
current and boundary heat current match precisely.  In sections 4 and 5,
we consider general Lovelock gravities and a class of Horndeski gravities
respectively. We derive the boundary heat currents and show that they
again match precisely with the corresponding radially conserved current.
We conclude the paper in section 6.

\section{Holographic heat current from Noether procedure}

\subsection{AdS planar black holes and a linear perturbation}

In this paper, we shall consider a general class of gravity theories in
$n$ dimensions, coupled to a set of matter fields including a Maxwell field.
We shall assume that the theory admits an AdS spacetime vacuum and that
the full action takes the form
\be
S=\int_{\cal M} d^n x \sqrt{-g} L(g_{\mu\nu}, A_\mu, \phi) +
\int_{\partial {\cal M}} d^{n-1}x
\sqrt{-h} (L_{\rm surf} + L_{\rm ct})\,,
\ee
where $A_\mu$ is the Maxwell field and $\phi$ denotes any additional matter
fields in the theory.  $L_{\rm surf}$ is the Gibbons-Hawking term or its
generalization, and $L_{\rm ct}$ denotes the holographic counterterms.
For simplicity, we shall assume that the Maxwell field is
minimally coupled to gravity, as in the case of the EMD theory.

  We shall be considering static background metrics of the form
\be
ds^2 = - \tilde f\, dt^2 +
\fft{dr^2}{f} + g^2\, r^2\, dx^i dx^i\,,\label{staticmet}
\ee
where $\tilde f$ and $f$ are functions of $r$, which at large $r$
approach the forms
\be
\tilde f(r) = g^2 r^2 + \cdots\,,\qquad f(r)= g^2 r^2 +\cdots\,,
\ee
where the ellipses denote terms of lower order in $r$. Thus the
metrics are asymptotic to AdS$_n$ spacetime with $R_{\mu\nu}=-(n-1) g^2\,
g_{\mu\nu}$. (Note that in this paper $g=1/\ell$ is the inverse of the
radius of the AdS, and it should not be confused with the determinant
of the metric.)  Also the coordinates $x_i$ have the same
(engineering) dimension as the time coordinate $t$, namely (Length)$^1$.
For the purpose of
calculating the heat current, we shall consider transverse traceless (TT)
metric perturbations for which the metric (\ref{staticmet}) takes the form
\be
ds^2 = - \tilde f\, dt^2 + \fft{dr^2}{f} +
g^2\, r^2\, dx^i dx^i + 2 \delta g_{tx_1}
\, dt dx_1 \,,\label{pertmet}
\ee
with $\delta g_{tx_1}$ being infinitesimal, and (in principle) depending on
all the coordinates.  However, for our purpose, we follow \cite{DC2} that the perturbation depends on $r$ only, except for a linear time dependence that is necessary for satisfying the ingoing boundary condition on the black hole horizon. The analogous strategy was developed for holographic DC currents \cite{DC1}.
Note that we did not include a perturbation $\delta g_{rx_1}(r)$, since it
can be can be removed by means of a coordinate transformation.

\subsection{Radially conserved current}

  Our derivation of a radially conserved current will closely follow the
procedure described by Wald \cite{wald1,wald2}.  Our starting point is a
bulk Lagrangian density
${\cal L}$, written in terms of a scale quantity $L_0$ as
\be
{\cal L} = \sqrt{-g} L_0 \,.
\ee
Under variations of the fields $\delta{\cal L}$ takes the form
\be
\delta {\cal L} = E.O.M. + \sqrt{-g} \, \nabla_\mu
J^\mu (\delta g_{\mu\nu}\,, \delta \phi \,,\delta A_\mu) \,.
\ee
where $E.O.M.$ denotes the equations of motion, and the additional
term collects
together the various total derivatives that do not contribute to the
equations of motion.

  In the Wald procedure, the two $(n-1)$-forms
$\Theta_{\sst{(n-1)}}$ and $J_{\sst{(n-1)}}$ are defined, by
\bea
\Theta_{\sst{(n-1)}} = * J_{\1} \,,\qquad J_{\sst{(n-1)}} =
\Theta_{\sst{(n-1)}} - \xi_\1 \cdot {* L_0} \,,
\eea
where $J_\1= J^\mu\, g_{\mu\nu}\, dx^\nu$, and
$\xi_\1 = \xi^\mu\, g_{\mu\nu}\, dx^\nu$ denotes the 1-form associated
with an  infinitesimal diffeormorphism $\delta x^\mu = \xi^\mu$. The term
$\xi_\1\cdot {*L}_0$ is the $(n-1)$ form obtained by contracting
$\xi$ onto the $n$-form ${*L}_0$.  The $(n-1)$ form $J_{\sst{(n-1)}}$ can
be written as
\be
J_{\sst{(n-1)}}= \Theta_{\sst{(n-1)}} - \xi_\1 \cdot {* L}_0=
d {* J}_\2 + E.O.M \,,  \label{nc}
\ee
where the gravity contribution to $J_\2$ is
\be
J_{\2\, {\rm gr}}^{\mu\nu} = 2 \fft{\partial L} {\partial R_{\mu\nu\rho\sigma}}
\nabla_\rho\xi_\sigma + 4 \xi_\rho \nabla_\sigma
\fft{\partial L} {\partial R_{\mu\nu\rho\sigma}}\,.\label{J2grav}
\ee
It is worth noting that for the Einstein-Hilbert term, we have
simply $J_{\2\,{\rm gr}}^{\mu\nu} = 2 \nabla^\mu\xi^\nu$.  Note also
that the contribution to $J_\2$ from the minimally-coupled Maxwell field is
\be
J_{\2 A}^{\mu\nu}=\xi^\rho A_\rho\, F^{\mu\nu}\,.\label{J2max}
\ee
On shell, the formula (\ref{nc}) turns out to be
\be
d {* J}_\2 = J_{\sst{(n-1)}} =  \Theta_{\sst{(n-1)}} - \xi_\1 \cdot {* L}_0
\,.
\ee
The Hodge dual of this equation is
\be
*d {* J}_\2 = {* J}_{\sst{(n-1)}} = {*\Theta}_{\sst{(n-1)}} -
{* (}\xi_\1 \cdot {* L}_0 )\,,
\ee
which gives
\be
\fft {1} {\sqrt{-g} } \partial_\nu(\sqrt{-g}  J_\2^{\mu\nu})  =
J_\1^\mu(\delta g_{\mu\nu}\,, \delta \phi \,,\delta A_\mu) +
L_0 \, \xi^\mu \,.
\ee

   If we consider a static metric, with $\xi$ taken to be the timelike
Killing vector $\xi^\mu\, \del_\mu = \del/\del t$, we have
\bea
\delta g_{\mu\nu} = \nabla_{\mu}\xi_\nu + \nabla_{\nu}\xi_\mu =0 \,, \ \
\delta A_\mu = \xi^\nu \partial_\nu A_\mu + (\partial_\mu \xi^\nu) A_\nu
= 0 \,, \ \
\delta \phi = \xi^\mu \nabla_\mu \phi  = 0 \,,
\eea
which implies $J_\1^\mu = 0$. Thus we have
\be
\fft {1} {\sqrt{-g}} \partial_\nu(\sqrt{-g} J_\2^{\mu\nu})  = L_0\, \xi^\mu \,.
\ee
The components of the right-hand side are all zero except in the
$t$ direction, and so in particular we have the radial
conservation for the components $J_\2^{\mu\nu}$,
\be
\fft {1} {\sqrt{-g}} \partial_r (\sqrt{-g}\, J_\2^{x_i\, r}) = 0 \,,
\ee
where $\mu=x_i$ represents coordinate index values in the directions
of the
spatial boundary metric, and $\nu=r$ denotes the radial index direction.
Thus we obtain a radially conserved current
\be
{\cal J}^{x_i} =  \sqrt{-g}\,  J_\2^{rx_i}\,.\label{gencurrent}
\ee
In particular, the gravity and Maxwell field
contributions to the 2-form
are given by (\ref{J2grav}) and (\ref{J2max}) respectively.

It is worth remarking that the bulk Noether current (\ref{gencurrent}) is
radially conserved
for all static or stationary solutions, where $\del/\del t$ is a
Killing vector.  For the static backgrounds we are focusing on in this
paper,  this Noether current vanishes identically for the static
background configuration
itself, and it gives rise to a non-trivial result for the metric
(\ref{pertmet}), where there is a TT perturbation.  The linear time
dependence in
$\delta g_{tx_1}$ discussed under (\ref{pertmet}) will not affect
that ${\cal J}^{x_1}$ is radially conserved at the linear level. In the
next subsection, we argue that the current (\ref{gencurrent}) associated
with the perturbation in (\ref{pertmet}) is precisely the bulk dual to
the holographic heat current of the boundary field theory of the AdS
spacetime.  It is worth emphasising that this method can be generalised to
include other Killing vectors, such as $\partial_ {x_i}$, in which case
the radially conserved current is related to the $x_i x_j$ component of
the boundary stress tensor,  and can be used to calculate the holographic
shear viscosity \cite{lllqtp}.

\subsection{Holographic heat current}

Turning on a small electric field $E_i$ and thermal gradient
$\nabla_i T$ will generate an electric current $J^i$ and thermal
current $Q^i = T^{ti} - \mu J^i$, where $T^{ab} $ is the boundary stress
tensor, $i$ denotes the spatial boundary directions and $\mu$ is the
chemical potential. To first order, the generalised Ohm law is given by
\begin{equation}
\left(
  \begin{array}{c}
    J \\
    Q \\
  \end{array}
\right)
=
\left(
  \begin{array}{cc}
    \sigma  & \alpha T  \\
    \bar \alpha T  & \kappa T \\
  \end{array}
  \right)
  \left(
  \begin{array}{c}
    E \\
    - (\nabla T)/T \\
  \end{array}
\right)           \,,
\end{equation}
where $\sigma$ is the electric conductivity, $\bar \kappa$ is the thermal
conductivity and $\alpha\,,\bar \alpha $ are thermoelectric conductivities.

The goal of this paper is to examine a variety of gravity theories,
and to establish that the radially-conserved bulk current
${\cal J}^{x_i}$ in (\ref{gencurrent}) matches to the thermal
current $Q^i$ on the boundary, namely
\be
{\cal J}^{x_i}\Big|_{\rm boundary} = Q^i\,.\label{equality}
\ee
Since ${\cal J}^{x_i}$ is radially conserved, one can read off
${\cal J}^{x_i}$ using the horizon data, which can be obtained even in
the case where an exact solution is not known.  It is clear that for the
static background (\ref{staticmet}) both the left and right-hand sides of
the equation (\ref{equality}) vanish, and hence in this case the equality holds
trivially.  Our goal therefore is to show that the equality (\ref{equality})
holds nontrivially, at the linearised level,
for the TT perturbation $\delta g_{tx_1}$ in (\ref{pertmet}).

In the existing literature, the construction of the radially-conserved
current ${\cal J}$ was carried out on the case by case basis. The first
example was given in \cite{DC2} for the EMD theory, where the equality
(\ref{equality}) was also established.  Analogous results were also
obtained for the Einstein-Gauss-Bonnet theory, in \cite{gsge}.

     Our construction of the radially-conserved current through the
Noether procedure
in section 2 makes it straightforward to construct the bulk current for a
variety of theories.
It is still necessary to obtain the corresponding boundary thermal current
and then to establish the equality (\ref{equality}).   In section 3, we
review this analysis for the Einstein-Maxwell-Axion (EMA) theory.
In sections 4, 5, we establish (\ref{equality}) for general Lovelock
gravities and also a class of Horndeski gravities.

\section{Holographic heat current in EMA theory}

\label{sec:EMA}

In this section, we review the analysis of the holographic currents in
the EMA theory.  This allows us to establish notation, and to
relate the abstract bulk Noether current in the previous section to
the boundary quantities. The Lagrangian is given by
\be
{\cal L} = \sqrt{-g} \Big( R -2\Lambda - \ft14 F^2 -
\ft12\sum_{i = 1}^{n-2} (\partial \chi_i)^2 \Big) \,,
\ee
where $ F = dA $.  The axions $\chi_i$'s are introduced to provide momentum
dissipation, so that one can obtain finite results for the DC conductivity
and the thermal heat conductivity. We shall consider the
static metric ansatz (\ref{staticmet}), together with
\be
A = a(r) dt \,,\qquad \chi_i= \lambda x_i\,.
\ee
Note that this is a simpler version of the more general EMD theory discussed
in \cite{DC2}.  Charged AdS black holes were constructed in \cite{EMDC}.
The focus of this section is to establish the equality (\ref{equality})
for this theory, rather than to obtain the explicit result for the
heat conductivity, which is a special case of \cite{DC2}.  It follows that
the axions do not play any significant role in our discussion.  They
however must be included in this set to obtain a finite conductivity,
since they are responsible for the momentum dissipation \cite{EMDC}.

We now consider a perturbation along one spatial direction, say $x_1$,
with $\delta g_{tx_1}$, $\delta A_{x_1}$ and $\delta \chi_1$ turned on.

  For simplicity, here and after, we consider only static backgrounds, even
though the construction of bulk currents in the previous section can be
applied to all stationary backgrounds with a time-like Killing vector.
There are two radially-conserved currents associated with the perturbation.
One is the electric current of the Maxwell field,
\be
J = \sqrt{-g} F^{rx_1} \,.
\ee
After imposing boundary conditions on the perturbation, one can evaluate
the electric current on the horizon and then obtain the corresponding
electric conductivity \cite{EMDC}.

   One can go one step further to study the holographic thermal conductivity,
since there exists an additional radially-conserved Noether current,
following the discussion of section 2. It is given by
\be
{\cal J}^{x_1} = \sqrt{-g} (2 \nabla ^r \xi ^{x_1} + a \,  F^{rx_1})
\,.\label{emajcurrent}
\ee
Note that the first term in the bracket comes from the contribution of
the Einstein-Hilbert term, and the second comes from the contribution
of the Maxwell field.  For the AdS planar black hole background with
the diagonal metric, it is clear that the Noether current
(\ref{emajcurrent}) vanishes identically.  However, for the aforementioned
perturbation the current (\ref{emajcurrent}) is non-vanishing, and can
be evaluated at the linear level.  We are interested in showing that
this current matches with the heat current on the boundary at
$r\rightarrow \infty$. The large-$r$ leading falloff at the asymptotic
boundary for ${\cal J}^{x_1}$ is given by
\be
{\cal J}^{x_1}_{\rm lin} = (g r)^{n-2}
\big(  \fft{r \delta g_{tx_1}' - 2 \delta g_{tx_1}}{ r} -
\mu\,F^{rx_1} \big) \,.
 \label{emajcurrent2}
\ee
Note that we have imposed the gauge condition $a(r)=0$ on the horizon and
so the value of $a(r)$ at infinity is the chemical potential,
$a(\infty) = \mu$.

We now turn our attention to the boundary conserved currents.  Including
the constant counterterm contribution , the boundary stress tensor is
given by
\be
T^{ab} = 2 (K h^{ab} - K^{ab}) - 2 (n-2) g\, h^{ab}\,,
\ee
where $h^{ab}$ is the induced boundary metric and $K^{ab}$ is the
extrinsic curvature of the boundary. (Since the boundary spacetime is flat,
there are no counterterms associated with boundary curvature. Note also
that $g=1/\ell$ here is the inverse of the AdS radius, and should not
be confused with the determinant of the metric.)

Specific to our perturbations, we have
\be
T^{tx_1}_{\rm lin} = \fft{r \delta g_{tx_1}' - 2 \delta g_{tx_1}}{g^3 r^4},
\ee
where a prime denotes the derivative respect to $r$. The boundary
electric current along the $x_1$ direction is given, to linear order, by
\be
J_{\rm BD}^e = -n_\mu F^{\mu x_1} = - \fft{1}{g r} F^{rx_1} \,.
\ee
The boundary heat current in the presence of the electric current is
defined as
\be
 Q = (g r)^{n+1} T^{tx_1} + (gr)^{n-1} \mu J_{\rm BD}^e\,.
\ee
At linear order in the perturbation, it is given by
\be
Q_{\rm lin}
= (gr)^{n-2} \big(  \fft{r \delta g_{tx_1}' - 2 \delta g_{tx_1} }{r} - \mu F^{rx_1}  \big) \,.
\ee
Thus we see that the radially-conserved Noether current (\ref{emajcurrent2})
matches precisely to the thermal current on the boundary.

\section{Holographic heat current in general Lovelock gravities}

   We begin this section by setting up our conventions for the Lovelock
Lagrangian \cite{ll},
and deriving the form of the radially conserved 2-form current.
We then go on to calculate the energy-momentum tensor in the boundary theory,
and to show that it is indeed related to the radially conserved 2-form current.
As was seen in previous section, holographic heat currents can be discussed
in the pure gravity sector without the inclusion of the Maxwell field.
However, it is necessary to include also a mechanism for momentum
dissipation, in order to obtain a finite conductivity.  In this section, we
shall employ free axions spanning
the AdS planar directions in order to generate the momentum dissipation
\cite{EMDC}, as discussed in section \ref{sec:EMA}. As we have seen in
section \ref{sec:EMA}, the axions give no contribution to
$J_\2^{\mu\nu}$. Furthermore, their overall contribution to the
metric background profile functions is to add a constant to
$f$ and $\tilde f$ in the large $r$ expansion, without altering the asymptotic
form, which remains $f\sim \tilde f \sim g^2 r^2$.  It follows that
they play no r\^ole in the matching of the bulk currents and boundary
stress tensor at asymptotic infinity.

\subsection{Bulk theory and conserved current}\label{noethersec}

The general bulk Lovelock action is given by
\be
S_{\rm{bulk}} = \int d^nx\, \sqrt{-g}\, L\,,\qquad
   L= \sum_{k\ge 0} a_{(k)}\, E_{(k)}\,,\label{bulklovelock}
\ee
where we define
\be
E_{(k)} = \fft{(2k)!}{2^k}\,
\delta^{\mu_1\cdots \mu_{2k}}_{\nu_1\cdots \nu_{2k}}\,
R^{\nu_1\nu_2}_{\mu_1\mu_2}\, R^{\nu_3\nu_4}_{\mu_3\mu_4} \cdots
  R^{\nu_{2k-1}\, \nu_{2k}}_{\mu_{2k-1}\, \mu_{2k}}\,.\label{Ekdef}
\ee
Note that our multi-index Kronecker delta symbol is defined to have
unit strength, so
\be
\delta^{\mu_1\cdots\mu_{2k}}_{\nu_1\cdots\nu_{2k}} =
  \delta^{[\mu_1} _{\nu_1}\, \delta^{\mu_2}_{\nu_2}\cdots
   \delta^{\mu_{2k}]}_{\nu_{2k}}\,,
\ee
where the square brackets denote conventional unit-strength
antisymmetrisations
(so, for example, $X^{[\mu_1\cdots \mu_p]} = X^{[[\mu_1\cdots\mu_p]]}$).  Note
that with our choice of normalisation for the Lovelock terms $E_{(k)}$, we have
\be
E_{(k)} = R^k +\cdots \,,
\ee
with unit coefficient for the purely Ricci-scalar term, where the
ellipses denote all terms involving one or more
uncontracted Ricci tensor or Riemann tensor.  (So, in particular, $E_{(0)}=1$,
$E_{(1)}=R$ and $E_{(2)}= R^2 - 4 R^{\mu\nu} R_{\mu\nu} +
  R^{\mu\nu\rho\sigma}\, R_{\mu\nu\rho\sigma}$, etc.)
The contribution to the Einstein equation from the $k$'th Lovelock term is
given by
\be
E^{(k)\, \mu}_\nu = -\fft{(2k+1)!}{2^{k+1}}\,
  \delta^{\mu \mu_1\cdots \mu_{2k}}_{\nu\nu_1\cdots \nu_{2k}}\,
  R^{\nu_1\nu_2}_{\mu_1\mu_2}\cdots
   R^{\nu_{2k-1}\,\nu_{2k}}_{\mu_{2k-1}\, \mu_{2k}}\,.\label{lovelockeom}
\ee
Calculating the conserved current for the $k$'th Lovelock term, using
(\ref{J2grav}), we find
\be
J^{(k)\mu\nu} = \fft{a_{(k)}\, k\, (2k)!}{2^{k-1}}\,
\delta^{\mu\nu\alpha_1\cdots \alpha_{2k-2}}_{
                      \rho\sigma\beta_1\cdots\beta_{2k-2}}\,
  R^{\beta_1\beta_2}_{\alpha_1\alpha_2}\cdots
   R^{\beta_{2k-3}\, \beta_{2k-2}}_{\alpha_{2k-3}\, \alpha_{2k-2}}\,
 \nabla^\rho \xi^\sigma\,.\label{J2Lovelock}
\ee
(Note that the second term in (\ref{J2grav}) does not contribute, by virtue
of the Bianchi identities for the Riemann tensor.)

 We are interested in particular in calculating the linearised contribution
to $J^{(k)\mu\nu}$ resulting from the metric perturbation
given in (\ref{pertmet}). We find
\bea
J^{(k)\mu\nu}_{\rm lin} &=& \fft{a_{(k)}\, k\,(k-1)\,  (2k)!}{2^{k-1}}\,
\delta^{\mu\nu\alpha_1\cdots \alpha_{2k-2}}_{
                      \rho\sigma\beta_1\cdots\beta_{2k-2}}\,
  \bar R^{\beta_1\beta_2}_{\alpha_1\alpha_2}\cdots
   \bar R^{\beta_{2k-5}\, \beta_{2k-4}}_{\alpha_{2k-5}\, \alpha_{2k-4}}\,
\delta R^{\beta_{2k-3}\, \beta_{2k-2}}_{\alpha_{2k-3}\, \alpha_{2k-2}}\,
 \bar\nabla^\rho \xi^\sigma \nn\\
&& +
  \fft{a_{(k)}\, k\, (2k)!}{2^{k-1}}\,
\delta^{\mu\nu\alpha_1\cdots \alpha_{2k-2}}_{
                      \rho\sigma\beta_1\cdots\beta_{2k-2}}\,
  \bar R^{\beta_1\beta_2}_{\alpha_1\alpha_2}\cdots
   \bar R^{\beta_{2k-3}\, \beta_{2k-2}}_{\alpha_{2k-3}\, \alpha_{2k-2}}\,
 \delta \nabla^\rho \xi^\sigma\,,\label{J2pert}
\eea
where the barred quantities are calculated in the background of the
unperturbed metric (\ref{staticmet}).

   We now wish to evaluate the linearised contribution $\delta J^{(k) r x_1}$
in the limit that $r$ goes to infinity, in order to compare it with the
heat current calculated in the boundary theory by evaluating the
$tx_1$ component of the boundary stress tensor.  We may evaluate
(\ref{J2pert}) in the large-$r$ limit by simply allowing the metric
functions $h$ and $f$ to take their asymptotic forms $\tilde f=f=g^2 r^2$. In
this limit the background metric is exactly AdS$_n$, and so in particular we
have
\be
\bar R^{\mu\nu}_{\rho\sigma}= -2 g^2\, \delta^{\mu\nu}_{\rho\sigma}\,,
\ee
implying that (\ref{J2pert}) becomes
\bea
J^{(k)\mu\nu}_{\rm lin}  &=&
\fft{12\,(-1)^k\, a_{(k)}\, k(k-1) \, (n-4)!\, g^{2k-4}}{(n-2k)!}\,
   \delta^{\mu\nu\alpha_1\alpha_2}_{\rho\sigma\beta_1\beta_2}\,
\delta R^{\beta_1\beta_2}_{\alpha_1\alpha_2}\, \bar\nabla^\rho\xi^\sigma
\nn\\
&& -\fft{2 k\,(-1)^k\,  a_{(k)}\, (n-2)!\, g^{2k-2}}{(n-2k)!}\,
\delta^{\mu\nu}_{\rho\sigma}\,
  \delta\nabla^\rho\xi^\sigma\,.
\eea
We also have $\bar\nabla^r\xi^t=-\bar\nabla^t \xi^r= g^2 r$, and after some
algebra we find that
\be
J^{(k)rx_1}_{\rm lin}  =
\fft{ (-1)^{k-1}\, a_{(k)}\, k\, (n-3)! \, g^{2k-2}}{
        (n-2k-1)!}\, \Big(\delta g_{tx_1}' -\fft2r\, \delta g_{tx_1}\Big)\,,
\label{Jrx1res}
\ee
where a prime denotes a derivative with respect to $r$.  The radially conserved current is given by ${\cal J}^{x_1} = \sqrt{-g} J^{rx_1}$, which vanishes on the background with the diagonal metric.  For the TT perturbation (\ref{pertmet}), the radially-conserved Noether current associated with $\delta g_{tx_1}$, evaluated at large $r$, is given by
\bea
{\cal J}^{x_1}_{\rm lin}
 &=& \sqrt{-\bar g} \sum_{k\ge1}  J^{(k) rx_1}_{\rm lin}  \cr
&=& (g r)^{n-2} \sum_{k\ge 1} \fft{ (-1)^{k-1}\, a_{(k)}\, k\, (n-3)! \, g^{2k-2}}{(n-2k-1)!}\, \Big(\delta g_{tx_1}' -\fft2r\, \delta g_{tx_1}\Big) \,.
\label{bkct}
\eea

\subsection{Surface terms for the Lovelock actions}

  Just as in ordinary Einstein gravity, in order to write the theory in
such a way that it has a well-defined Hamiltonian formulation, it is
necessary to add a surface term to the action that removes the
$\nabla \delta g_{\mu\nu}$ terms arising in the variational principle.
For Einstein gravity, the necessary boundary term, which was derived by
York \cite{york} and by Gibbons and Hawking \cite{Gibbons:1976ue}, involves
the trace of the second
fundamental form on the boundary.  A general discussion of the analogous
surface terms for general gravity theories was given in \cite{deruelle}.  Their
discussion made use of an auxiliary field formulation, and after
restating it in terms of just the original metric formulation, it may be
expressed as follows.  We begin by defining a surface action
$\widetilde S_{\rm{surf}}$ as
\be
S_{\rm{surf}} = 4 \int \sqrt{-h}\, d^{n-1}x\,
   \Psi^{\mu\nu} K_{\mu\nu}
\,,
\ee
where $K_{\mu_\nu}= h_\mu{}^\rho\, \nabla_\rho \, n_\nu$ is the second
fundamental form, $h_{\mu \nu}=g_{\mu\nu}- n_\mu n_\nu$ and $n^\mu$ is
the unit normal vector on the boundary (spacelike, with
$n=f^{1/2}\, \del/\del r$ in our case). Eventually, the auxiliary
field $\Psi^{\mu\nu}$ is solved for, and is given by
\be
\Psi^{\mu\nu}= \fft{\del L}{\del R_{\mu\rho\nu\sigma}} \, n_\rho\,n_\sigma
\,.\label{Psidef}
\ee
Varying $S_{\rm{surf}}$ with respect to the boundary metric to
obtain the boundary contributions that will subtract those coming from the
integrations by parts for the bulk action, one leaves $\Psi^{\mu\nu}$
unvaried and only then makes the substitution (\ref{Psidef}).

   One can instead construct $S_{\rm{surf}}$ directly as follows.  We
begin by defining
\be
\widetilde S_{\rm{surf}} = 4 \int \sqrt{-h}\, d^{n-1}x\,
  \fft{\del L}{\del R^{\mu\nu}_{\rho\sigma}}\, K^\mu_\rho\, n^\nu\, n_\sigma\,.
\label{Stildedef}
\ee
Using the Gauss-Codacci equation
\be
R^{ab}_{cd}= {\cal R}^{ab}_{cd} - 2 K^a_{[c}\, K^b_{d]}\,,\label{GaussC}
\ee
we substitute into the expression for $\widetilde S_{\rm{surf}}$.  We may
then calculate the desired surface term $S_{\rm{surf}}$ by means of
the equation
\be
K_{ab}\, \fft{\del S_{\rm{surf}}}{\del K_{ab}} =
   \widetilde S_{\rm{surf}}\,.\label{SfromStilde}
\ee
By this means, one is compensating for the fact that (\ref{Stildedef})
contains higher powers of $K$, which, when varied, would give too large
a contribution in the variation.  We can express the solution to
(\ref{SfromStilde}) in the integral form
\be
S({\cal R}^{ab}_{cd}, K^a_b)_{\rm{surf}}
 = \int_0^1 \widetilde S({\cal R}^{ab}_{cd}, u\, K^a_b)_{\rm{surf}}\,
\fft{du}{u}\,.
\label{intform}
\ee

   The result of solving (\ref{SfromStilde}) is very easy to state in the
case where we consider the $k$th Lovelock Lagrangian and where we also
take the boundary to be flat (as in our discussion in this paper).  We then
have simply
\be
 S^{(k)}_{\rm{surf}} = \fft1{2k-1}\, \widetilde S^{(k)}_{\rm{surf}}\,,
\ee
and so
\bea
S^{(k)}_{\rm{surf}} &=& \int \sqrt{-h} \,
d^{n-1}x\,  L^{(k)}_{\rm{surf}}\,,\nn\\
L^{(k)}_{\rm{surf}} &=&\fft{(-1)^{k-1}\, (2k)!\, a_{(k)}}{2k-1}\,
    \delta^{a_1\cdots a_{2k-1}}_{b_1\cdots b_{2k-1}}\,
   K^{b_1}_{a_1}\, K^{b_2}_{a_2}\cdots K^{b_{2k-1}}_{a_{2k-1}}\,.
\label{surfacepureK}
\eea

   If we consider the case where the boundary metric is curved, then from
(\ref{intform}) we have:
\bea
L^{(k)}_{\rm{surf}} &=& \fft{(2k)!\, a_{(k)}}{2^{k-1}}\,
  \delta^{a_1\cdots a^{2k-1}}_{b_1\cdots b_{2k-1}}\times\\
&&
 \int_0^1 du\, ({\cal R}^{b_1 b_2}_{a_1 a_2} - 2 u^2 \, K^{b_1}_{a_1}
   K^{b_2}_{a_2})
   \cdots ({\cal R}^{b_{2k-3} \, b_{2k-2}}_{a_{2k-3}\, a_{2k-2}}
  -2 u^2\, K^{b_{2k-3}}_{a_{2k-3}} K^{b_{2k-2}}_{a_{2k-2}})
   \, K^{b_{2k-1}}_{a_{2k1-}}\,.\nn
\eea
Written explicitly, we have
\bea
L^{(k)}_{\rm{surf}} &=& \fft{(2k)!\, a_{(k)}}{2^{k-1}}\,
\sum_{\ell=0}^{k-1}\, \fft{(k-1)!\, (-2)^\ell}{(k-\ell-1)!\,
                         \ell!\, (2\ell+1)}\,
\delta^{a_1\cdots a_{2k-2\ell -2}\, a_{2k-2\ell -1}\cdots a_{2k-1}}_
{b_1\cdots b_{2k-2\ell -2}\, b_{2k-2\ell -1}\cdots b_{2k-1}}\times\nn\\
&&\qquad\qquad \qquad {\cal R}^{b_1 b_2}_{a_1 a_2}\cdots
{\cal R}^{b_{2k-2\ell-3}\,b_{2k-2\ell-2}}_{a_{2k-2\ell-3}\,a_{2k-2\ell-2}} \,
K^{b_{2k-2\ell-1}}_{a_{2k-2\ell-1}} \cdots K^{b_{2k-1}}_{a_{2k-1}}\,.
\eea
Evaluating this for $k=1$ gives the standard Gibbons-Hawking boundary term
for the bulk gravitational Lagrangian $a_{(1)} R$:
\be
L^{(1)}_{\rm{surf}}  = 2 a_{(1)}\, K\,,
\ee
where $K=K^a_a$, while for $k=2$ we obtain
\be
L^{(2)}_{\rm{surf}} = - 4 a_{(2)}\, \Big[ 2 {\cal G}^a_b\, K^b_a +
        \ft13( K^3 - 3K\, K^a_b\, K^b_a + 2 K^a_b\, K^b_c\, K^c_a)\Big]\,,
\ee
where ${\cal G}_{ab}= {\cal R}_{ab}- \ft12 {\cal R}\, h_{ab}$ is
the boundary Einstein tensor.  This result
 agrees with the surface term for Gauss-Bonnet gravity given in
\cite{myersgs,jmliugs}.

\subsection{Flat-boundary counterterms for Lovelock actions}

  In the previous subsection, we derived explicit general formulae for the
surface terms for all Lovelock actions, with curved as well as flat boundaries.
In fact, for the purposes of this paper, we are interested in the
restricted results for the case where the boundary metric is flat, in which
case the simple expressions given by (\ref{surfacepureK}) are sufficient.
In this subsection, we shall calculate the counterterms for all Lovelock
actions.  Here, however, we shall restrict ourselves from the outset to the
case where the boundary metric is flat, since otherwise the calculations
would become to unwieldy, and we do not in any case need the counterterms
involving boundary curvature for the purposes of this paper.

   The counterterm action $S_{\rm ct}$
that we seek can be determined by requiring that the leading-order
power-law $r$ divergence in the total action
$S_{\rm tot} = S_{\rm{bulk}} + S_{\rm{surf}} + S_{\rm ct}$ be cancelled.
This will ensure that the total action is finite for AdS$_n$ itself
 (corresponding to the metric (\ref{staticmet}) with
$\tilde f = f=g^2\, r^2$), and
this uniquely determines the general expression for the counterterms for a
flat boundary metric.  Note that in this AdS$_n$ background we have
$\sqrt{-g}= (gr)^{n-2}$ and $\sqrt{-h}= (gr)^{n-1}$, together with
$R^{\mu\nu}_{\rho\sigma}=-2 g^2\, \delta^{\mu\nu}_{\rho\sigma}$ and
$K^a_b= g\, \delta^a_b$.

   Using (\ref{lovelockeom}), we can see that the equations of motion for the
Lovelock theory with bulk action given by (\ref{bulklovelock}) imply, for
the pure AdS$_n$ background, that
\be
-\ft12 \sum_{k\ge 0} \fft{(n-1)!}{(n-2k-1)!}\, (-g^2)^k\, a_{(k)}=0\,.
\ee
One may think of this equation as determining the ``bare'' cosmological
constant $-\ft12 a_{(0)}$ in terms of the AdS$_n$ scale-size parameter $g$,
for given values of the higher Lovelock couplings $a_{(k)}$ with $k\ge 1$.

   The on-shell bulk action in the AdS$_n$ background, integrated
out to a radius $\bar r$, is given by
\be
S_{\rm bulk}= \sum_{k\ge 0}\, S^{(k)}_{\rm bulk} =
   \int d^{n-1}x\, \sum_{k\ge 0}\, a_{(k)}\, \int^{\bar r} \, dr\, \sqrt{-g}\,
  E^{(k)}\,,\label{bulkonshell}
\ee
where on-shell we have, from (\ref{Ekdef}), that
\be
E^{(k)} =\fft{n!\, (-g^2)^k}{(n-2k)!}\,.
\ee
The surface terms, which can be determined from (\ref{surfacepureK}) with
$K^a_b=g\, \delta^a_b$, give a total on-shell surface action
\be
S_{\rm surf} = \int d^{n-1}x\, \sqrt{-h}\,
            \sum_{k\ge 1} L^{(k)}_{\rm surf}\,,\qquad
L^{(k)}_{\rm surf} = \fft{2k\,  g\, (n-1)!\, (-g^2)^{k-1}\, a_{(k)}}{
               (2k-1)\, (n-2k)!}\,.
\ee
Evaluating $S_{\rm surf}$ at $r=\bar r$, and combining it with the contribution
coming from (\ref{bulkonshell}), we find that the power-law $\bar r$
divergence, arising at $\bar r^{n-1}$ order, is cancelled provided we add
a counterterm action
\be
S_{\rm ct} = \int d^{n-1}x\, \sqrt{-h}\, \sum_{k\ge 1}
   \fft{(n-2)!\,2k\, (-1)^k\,
     a_{(k)}}{(2k-1)\, (n-2k-1)!}\, g^{2k-1}\,.\label{Sct}
\ee

\subsection{Boundary energy-momentum tensor}

Having obtained the complete action
\be
S_{\rm tot} = S + S_{\rm ct}\,,\qquad S=S_{\rm bulk} + S_{\rm surf}\,,
\ee
we are now in a position to calculate the boundary energy-momentum tensor
\be
T^{ab} = \fft{2}{\sqrt{-h}} \fft{\delta S_{\rm tot}}{\delta h_{ab}}=
2\pi^{ab} + T_{\rm ct}^{ab}\,,\label{Tab}
\ee
where
\bea
T^{ab}_{\rm ct} &=& \fft{2}{\sqrt{-h}}\,
\fft{\delta S_{\rm ct}}{\delta h_{ab}}\,,\nn\\
&=& \sum_{k\ge 1} \fft{(n-2)!\, 2k\, (-1)^k\, a_{(k)} g^{2k-1}}{
    (2k-1)\, (n-2k-1)!}\, h^{ab}\,.\label{Tabct}
\eea
(In obtaining the last line, we made use of the expression (\ref{Sct}).)

    The canonical momentum $\pi^{ab}$ is given by
\be
\pi^{ab}\equiv \fft{2}{\sqrt{-h}} \fft{\delta S}{\delta h^{ab}}\,,
\ee
evaluated on the boundary. Although we have explicitly obtained the
surface action in the previous subsection 4.2, the direct calculation
of $\pi^{ab}$ from this would be somewhat involved.  For a theory
involving only second derivatives, it is more convenient to make use of the
observation in \cite{deBoer:1999tgo}, which implies that
\be
\pi^{ab} = \fft{1}{\sqrt{-h}} \fft{\delta S}{\delta \dot h_{ab}} =
\fft{1}{2\sqrt{-h}} \fft{\delta S}{\delta K_{ab}} =
\fft{1}{2} \fft{\partial L}{\partial K_{ab}},
\ee
where $L$ is defined by $S=\int d^n x \sqrt{-g} L$.
The expression of $L$ in the ADM decomposition can best be stated in terms
of the Lagendre transformation
\be
K_{ab} \fft{\partial L}{\partial K_{ab}} - L =
H \equiv - L_{\rm bulk} (R^{ab}_{cd})\,,\label{Legendre}
\ee
where the expression for the Hamiltonian $H$ in the last equality was
demonstrated, for the Lovelock theories, in \cite{Teitelboim:1987zz}.
Here $R^{ab}_{cd}$ denotes the restriction of $R^{\mu\nu}_{\rho\sigma}$
to its components purely in the boundary directions,
which are then expressed in
terms of ${\cal R}^{ab}_{cd}$ and $K^a_b$ by using the
Gauss-Codacci equations, as in (\ref{GaussC}).  From the Legendre
transformation (\ref{Legendre}) it follows that
\be
K_{cd}\, \fft{\del \pi^{ab}}{\del K_{cd}} = \tilde \pi^{ab}\,,\qquad
\hbox{where}\quad \tilde \pi^{ab} = \fft12\, \fft{\del H}{\del K_{ab}}\,.
\label{pifrompitilde}
\ee
The key point here is that $H({\cal R}^{ab}_{cd},K^a_b)$ is easily
calculated, as in the final equality in (\ref{Legendre}), and hence
$\tilde \pi^{ab}$ is easily obtained.  Specifically, the contribution
$H^{(k)}$ for the $k$th Lovelock Lagrangian gives
\be
H^{(k)}= -\fft{a_{(k)}\, (2k)!}{2^k}\,
\delta^{a_1\cdots a_{2k}}_{b_1\cdots b_{2k}}\,
({\cal R}^{b_1 b_2}_{a_1 a_2} - 2 K^{b_1}_{a_1} \, K^{b_2}_{a_2})\cdots
({\cal R}^{b_{2k-1}\, b_{2k}}_{a_{2k-1}\, a_{2k}} - 2K^{b_{2k-1}}_{a_{2k-1}}\,
  K^{b_{2k}}_{a_{2k}})\,.\label{Hkgen}
\ee
Then,
(\ref{pifrompitilde}) can be used in order to calculate $\pi^{ab}$
from $\tilde\pi^{ab}$.

   In our case the metric on the boundary is flat, and so ${\cal R}^{ab}_{cd}$
vanishes.  This means that for the $k$'th Lovelock
Lagrangian, $H^{(k)}$ given in (\ref{Hkgen}) reduces to
\be
H^{(k)}= - (-1)^k\, a_{(k)}\, (2k)!\,
\delta^{a_1\cdots a_{2k}}_{b_1\cdots b_{2k}}\,
K^{b_1}_{a_1} \cdots
  K^{b_{2k}}_{a_{2k}}\,.\label{Hkflat}
\ee
It is homogeneous
of degree $2k$ in $K^a_b$, and hence
$\tilde\pi_{(k)}^{ab}$ is homogeneous of degree $(2k-1)$ in $K^a_b$.  It then
follows from (\ref{pifrompitilde}) that
\be
\pi^{ab}_{(k)} = \fft1{(2k-1)}\, \tilde\pi^{ab}_{(k)}=\fft1{2(2k-1)}\,
   \fft{\del H_{(k)}}{\del K_{ab}}\,.
\ee
 From (\ref{Hkflat}) we therefore have
\be
\pi^a_{(k)\, b} = - \fft{(-1)^k\, k\,a_{(k)}\,  (2k)!}{(2k-1)}\,
   \delta^{a a_1\cdots a_{2k-1}}_{b b_1\cdots b_{2k-1}}\,
    K^{b_1}_{a_1}\cdots K^{b_{2k-1}}_{a_{2k-1}}\,.\label{piab}
\ee

  Evaluating first the background value of $\pi^a_{(k)\, b}$, which
we shall denote by $\bar \pi^a_{(k)\, b}$, we find
\be
\bar \pi^a_{(k)\, b} = -
\fft{(-1)^k\, k\,a_{(k)}\,
 (n-2)!\, g^{2k-1}}{(2k-1)\, (n-2k-1)!}\, \delta^a_b\,.
\label{barpiab}
\ee
Varying (\ref{piab}) to get the linearised perturbation in $\pi^a_{(k)\, b}$
we find
\be
\delta \pi^a_{(k)\, b} =-\fft{(-1)^k\, k\, a_{(k)}\, (n-3)!}{(n-2k-1)!}\,
  (\delta K^c_c\, \delta^a_b - \delta K^a_b)\,.
\ee
Finally, we obtain the linearised perturbation in $\delta\pi_{(k)}^{ab}$ as
\be
 \delta\pi_{(k)}^{ab} =\delta\pi^a_{(k)\, c} \, \bar h^{bc} +
     \bar\pi^a_{(k)\, c}\, \delta h^{bc}\,,\label{varpiab}
\ee
where $\bar h^{ab}$ is the background value of $h^{ab}$ and $\delta h^{ab}$
is its linearised perturbation.

    From (\ref{Tab}), (\ref{Tabct}), (\ref{barpiab}) and (\ref{varpiab})
we see that in calculating the linearised boundary stress tensor
$T^{ab}_{\rm lin}$,
the
$\bar\pi^a_{(k)\, c}\, \delta h^{bc}$ term in (\ref{varpiab}) is
precisely cancelled by the linearised counterterm contribution
$T^{ab}_{\rm ct,\, lin}$, and so we arrive at the final result
\be
\delta T_{(k),\, {\rm lin}}^{ab} = -\fft{ (-1)^k\, 2k\, a_{(k)}\, (n-3)!\, g^{2k-2}}{
   (n-2k-1)!}\, (\delta K^c_c\, \delta^a_b -\delta K^a_b)\,.\label{varTab}
\ee

   We are interested in comparing the expression for the $tx_1$ component
of the linearised boundary stress tensor with the expression for the
bulk Noether current $\delta J^{(k)\, r x_1}$ that we calculated in
section (\ref{noethersec}).  For our perturbed metric (\ref{pertmet}) we
have, at the linearised order, $\delta K^c_c=0$, and $\delta K^t{}_{x_1}=
   -1/(2 g r)\, (\delta g_{tx_1}' - 2 \delta g_{tx_1}/r)$ and so
\bea
T_{(k),\, {\rm lin}}^{tx_1}&=&
\fft{(-1)^{k-1}\, k\, a_{(k)}\, (n-3)!\, g^{2k-2}}{
(n-2k-1)!\, (g r)^3}\, \Big(\delta g_{tx_1}' - \fft2{r}\, \delta g_{t x_1}
\Big)\,,\\
&=& \fft1{(g r)^{n+1}}\, {\cal J}^{(k)\, x_1}_{\rm lin}\,,
\eea
where ${\cal J}^{(k)\, x_1}$ is the $k$'th term of the bulk
radially-conserved Noether current obtained in eqn (\ref{bkct}). Thus
we have established the desired correspondence between the
radially conserved bulk current and the boundary heat current.

\section{Holographic heat current in Horndeski gravities}

In this section, we consider Einstein-Horndeski gravities, which are a
class of higher-derivative theories involving non-minimally coupled scalar
axion fields.  For simplicity, we shall include in the
Lagrangian just the simplest non-trivial Horndeski term, namely
\cite{Horndeski:1974wa}
\be
S = \int d^nx \sqrt{-g} \big( R - 2 \Lambda - \ft12 \alpha \chi^\mu\chi_\mu + \ft12 \gamma  G_{\mu\nu} \chi^\mu\chi^\nu \big)\,,
\ee
where $\chi$ is a scalar field, $\chi^\mu \equiv \del^\mu \chi$,
and $G_{\mu\nu} = R_{\mu\nu} - \ft12 R g_{\mu\nu}$ is the Einstein tensor.
Static AdS black hole solutions with $\chi=\chi(r)$ were constructed in
\cite{ac,aco,Rinaldi:2012vy,Babichev:2013cya}. The thermodynamics of these
solutions has been analysed in \cite{Feng:2015oea,Feng:2015wvb}.
(See also \cite{caceres}.)

The AdS planar black hole solution is particularly simple, and is given by
\bea
ds^2 &=& - f dt^2 + \fft{dr^2}{f} +g^2 r^2 dx^i dx^i \,, \qquad \chi= \chi(r) \,, \cr
f& =& g^2 r^2 - \fft{\mu}{r^{n-3}} \,, \qquad \chi' = \sqrt{\fft{\beta}{f}}\,.
\eea
Here a prime denotes a derivative with respect to $r$. The parameters
$(\beta, g)$ are related to the coupling constants by
\be
\Lambda = - \ft12 (n-1)(n-2)  (1 + \ft12 \beta\gamma) g^2 \,,
\qquad \alpha =  \ft12 (n-1)(n-2) g^2 \gamma \,.
\ee
The solution contains only one non-trivial integration constant $\mu$,
proportional to the mass of the black hole.

It follows from (\ref{J2grav}) that the bulk current is given by
\bea
J^{\mu\nu} &=&
 2 \nabla^\mu\xi^\nu + \, \gamma  \Big[ - \ft12 (\partial \chi)^2 \nabla^\mu \xi^\nu + \nabla^\sigma \chi \nabla^{[\mu}\chi \nabla_\sigma \xi^{\nu]} + \xi^{[\nu}\nabla^{\mu]} (\partial \chi)^2 \cr
 && \qquad - \xi^{[\nu} \nabla_\sigma (\nabla^{\mu]} \chi \nabla^\sigma \chi) - \xi^\sigma \nabla^{[\mu} ( \nabla^{\nu]} \chi \nabla_\sigma \chi  )  \Big] \,.
\eea
Considering a linearised TT metric perturbation (\ref{pertmet}) and taking
the Killing vector $\xi$ to be $\partial_t$, we find that
the $rx_1$ component of the 2-form $J_\2$ at linear order,  at large $r$,
is given by
\be
J^{rx_1}_{\rm lin} = \Big(1 + \fft{\beta \gamma}{4}\Big) \,
\fft{r \delta g_{tx_1}' - 2 \delta g_{tx_1}}{r} \,.
\ee
Thus the radially-conserved bulk current is
\be
\delta {\cal J}^{x_1} = \sqrt{-\bar g}\, \delta J^{tx_1} = \sqrt {-\bar g}
\Big(1 + \fft{\beta \gamma}{4}\Big) \,
\fft{r \delta g_{tx_1}' - 2 \delta g_{tx_1}}{r} \,.\label{hornJ1}
\ee
As in the case of the Lovelock gravities we discussed previously, the goal now
will be to establish that the  above bulk current
is the holographic dual to the boundary heat current, i.e. that it
matches to the boundary stress tensor.

The generalized Gibbons-Hawking term is
\be
S_{\rm surf} =  4 \int dx^{n-1} \sqrt{-h}\, \fft{\partial L}{\partial R^{\mu\nu}_{\rho\sigma}} \,K^\mu_\rho n^\mu n_\sigma = 2\int dx^{n-1} \sqrt{-h}\, K\,.
\ee
In other words, the Horndeski term does not modify the surface term. As in
the previous section on Lovelock theories, we can use the
Gauss-Codacci relation
to express the bulk curvature in terms of the intrinsic boundary
curvature ${\cal R}^{ab}_{cd}$ and the extrinsic curvature $K^a_b$:
\bea
G_{\mu\nu} n^\mu n^\nu  &=& - \ft12 h^{\mu\nu} h^{\rho\sigma} R_{\mu\rho\nu\sigma} = - \ft12 ({\cal R} - K^2 + K_{ab}^2)\,,\nn\\
R &=& {\cal R } + K^2 - K_{ab}^2 + 2 \nabla_\alpha(n^\beta \nabla_\beta n^\alpha - n^\alpha \nabla_\beta n^\beta) \,.
\eea

For the AdS planar black hole in Horndeski gravity we considered above,
the scalar axion $\chi$ depends only on the coordinate $r$, and the
Horndeski term can be thus written as
\bea
G^{\mu\nu} \chi_\mu\chi_\nu  = G_{\mu\nu} \chi_\rho \chi_\sigma n^\mu n^\nu n^\rho n^\sigma  = - \ft12 ({\cal R} - K^2 + K_{ab}^2) \chi^r \chi_r \,.
\eea
Indeed, this result is consistent with the earlier observation that there is
no Gibbons-Hawking type surface contribution associated with the Horndeski term.

The total bulk action together with Gibbons-Hawking surface term, expressed
in terms of the induced metric and exterior curvature $K_{\mu\nu}$, is given by
\be
S_{\rm bulk}+S_{\rm surf} = \int d^nx \sqrt{-g} \Big[(1 + \ft14 \gamma \,
\chi^r \chi_r)(K^2 - K_{ab}^2) - 2 \Lambda - \fft \alpha 2 \chi^\mu\chi_\mu \Big] \,.
\ee
Note that the intrinsic curvature contribution vanishes, i.e.~${\cal R} = 0$,
since
the boundary of the AdS planar black hole we consider is flat. The corresponding Hamiltonian is then given by
\be
{\cal H } = K_{ab} \fft{\partial L} {K_{ab}} - L =
(1 + \ft14 \gamma\, \chi^r \chi_r)(K^2 - K_{ab}^2)\,,
\ee
from which we obtain the canonical momentum
\be
\pi^{ab} = \fft 12 \fft{\partial {\cal H}}{\partial K_{ab}} = (1 + \ft14
\gamma\,\chi^r \chi_r)(K h^{ab} - K^{ab})\,.
\ee

The counterterm for this theory can be obtained by using the same strategy
as in our discussion of the Lovelock gravities.  It was in fact obtained
in \cite{Feng:2015oea}, and is given by
\be
S_{\rm ct} = - \int dx^{n-1} \sqrt h \, 2 (n-2) ( 1 + \ft14 \beta \gamma)
g \,.
\ee
The contribution to the boundary stress tensor from the counterterm is
\be
T^{ab}_{\rm ct} = \fft{2}{\sqrt{-h}} \fft{\delta S_{ct}}{\delta h_{ab}}
= - 2 (n-2) (1 + \ft14 \beta \gamma) g h^{ab}\,.
\ee
The full boundary stress tensor is therefore given by
\be
 T^{ab} = 2 \pi^{ab} + T^{ab}_{\rm ct} = 2
(1 + \ft14 \gamma \, \chi^r \chi_r)(g h^{ab} - K^{ab}) \,.
\ee
The ${tx_1}$ component of the stress tensor, to linear order in
the perturbation, is then given by
\be
T^{tx_1}_{\rm lin} = (1 + \ft14 \beta\gamma) \,
\fft{r \delta g_{tx_1}' - 2 g_{tx_1}}{g^3 r^4} \,.
\ee
The corresponding heat current is
\be
{\cal Q}^{x_1} = (gr)^{n+1} T^{tx_1}_{\rm lin} =
(1 + \ft14 \beta \gamma) g^{n-2} r^{n-3} \,
(r \delta g_{tx_1}' - 2 \delta g_{tx_1}) \,.
\ee
It can easily be seen that it matches precisely on the boundary with the
radially-conserved Noether current $\delta{\cal J}^{x_1}$ given in
(\ref{hornJ1}).  Note that although we established the matching of the bulk
Noether current and boundary heat current using the explicit AdS planar
black hole solution, the matching works even for more general black holes
with additional matter fields, as long as the asymptotic metric functions
take the form $\tilde f\sim f \sim g^2 r^2$ at large $r$.

Einstein-Horndeski gravities with multiple Horndeski axions can also
admit a different class of black hole solutions, in which the axions play
the direct role of providing the momentum dissipation. The Lagrangian is
given by
\be
\fft{\cal L}{\sqrt{-g}} = R-2\Lambda - \ft14 F^2 -
\ft{1}{2}(\alpha g^{\mu\nu}-\gamma G^{\mu\nu})\, \sum_{i=1}^N
 \del_\mu\chi_i\, \del_\nu\chi_i \,.\label{lag3}
\ee
The ansatz for the AdS planar black holes is
\bea
ds^2_n &=& -h(r) dt^2 + \fft{dr^2}{f(r)} + r^2 (dx_1^2 + \cdots + dx_{n-2}^2)\,,\cr
A &=& a(r)\, dt\,,\qquad \chi_i=
\left\{
  \begin{array}{ll}
    \lambda x_i, &\qquad i=1,2,\ldots,n-2\,, \\
    0, &\qquad i\ge n-1\,.
  \end{array}
\right.
\eea
The four-dimensional solution was constructed in \cite{Jiang:2017imk}, and
the higher-dimensional generalizations were given in \cite{Feng:2017jub}.
The holographic DC conductivities were analysed in \cite{Jiang:2017imk}.
 The four-dimensional radially-conserved bulk current associated with the
holographic heat current was given in \cite{Baggioli:2017ojd}, although
its match with the boundary stress tensor was not demonstrated.

It follows from (\ref{J2grav}) and (\ref{gencurrent}) that the
radially-conserved Noether current at large $r$ is given by
\be
{\cal J}^{x_1}_{\rm lin} = \sqrt{-\bar g}\, (r\delta g_{tx_1}' - 2 g_{t x_1})
\Big(\fft1{r} -
\fft{n-4}{4r^3} \gamma \lambda^2\Big)\,.
\ee
It is evident that the Horndeski term contributes a sub-leading order
relative to the contribution from the Einstein-Hilbert term on the boundary,
 and hence it can be neglected.  The same is true for the boundary stress
tensor, as can be seen on the grounds of dimensional analysis; namely,
the quantity $\gamma \lambda^2$ has dimension of length squared.  It
then follows
that the matching (\ref{equality}) holds straightforwardly.

\section{Conclusions}

A radially conserved current in the bulk that matches the boundary heat
current is a key ingredient in the holographic study of thermal
conductivity
and related transport coefficients.  It allows one to read off the
relevant transport properties directly and analytically from the
black hole horizon data of the solution in the bulk theory. In this paper,
we employed a Noether procedure, closely related to that used
by Wald \cite{wald1,wald2}, and derived a formula (\ref{gencurrent}) for
a radially conserved bulk current for a general class of gravity theories,
in the case of AdS planar black hole backgrounds (\ref{staticmet}) with a
TT perturbation of the form (\ref{pertmet}).  We showed that the formula
(\ref{gencurrent}) reproduced known results in literature, including the
EMD theory \cite{DC2}, Einstein-Gauss-Bonnet gravity \cite{gsge} and also
four-dimensional Eintein-Horndeski gravity \cite{Baggioli:2017ojd}.

   In order to demonstrate that the Noether current (\ref{gencurrent}) is
indeed the holographic bulk dual to the heat current, it is necessary to
show that it matches the boundary heat current derived from the
boundary stress tensor.  We focused our discussion on the general
class of all Lovelock higher-derivative gravities.  We derived the
generalized Gibbons-Hawking surface term for the general Lovelock gravities,
and we constructed the boundary counterterms for the flat boundaries that
arise for
AdS planar black holes.  This enabled us to construct the full boundary
stress tensor and to derive the heat current associated with TT
perturbations.  We showed that the bulk Noether current and the
boundary heat current
match precisely.  We also performed an analogous analysis for a class of
Horndeski gravities in general dimensions, and showed again that
the Noether current (\ref{gencurrent}) is the relevant radially-conserved
current that describes the holographic heat current.

     Our discussion was concerned only with ghost-free theories such as
Lovelock and Horndeski gravities. Even though the total number of
derivatives in each term in the equations of motion can exceed two this
occurs only because of the nonlinearities; the linearized equations of
motion around any background involve only two derivatives at most.
This makes it easier to construct the Hamiltonian and to
derive the boundary stress tensor.  Indeed, as we have shown in these cases
the Noether current (\ref{gencurrent}) matches precisely the heat
current derived from the boundary stress tensor.

     The Noether current (\ref{gencurrent}) on the other hand is radially
conserved for more general classes of gravity theories that may have ghost
excitations as well.  We expect that it is still the relevant holographic bulk
dual to the heat current even in these more general situations, but this
remains to be investigated in detail.

\section*{Acknowledgement}

We thank Aristos Donos, Luis Fernando, Jerome Gauntlett and Tom Griffin 
for useful comments on our paper. C.N.P.~is grateful to the Center for 
Advanced Quantum Studies and the physics department at Beijing Normal 
University for hospitality during the course of this work.
H-S.L.~is supported in part by NSFC grants No.~11305140, No.~11375153,
No.~11475148 and No.~11675144.
H.L.~is supported in part by NSFC grants No.~11475024, No.~11175269 and
No.~11235003. C.N.P.~is supported in part by DOE grant DE-FG02-13ER42020.

\end{document}